\documentclass[12pt,notitlepage,a4paper]{article}
\pdfoutput=1
\usepackage{color,graphicx}
\usepackage{amssymb}
\usepackage{delarray,amsmath,bbm}
\usepackage[latin1]{inputenc}
\usepackage[american]{babel}
\usepackage{cite}
\usepackage{subfigure}
\usepackage{epsfig}

\begin{document}

\begin{center}
{\Large \bf  Oscillating Shells in Anti-de Sitter Space$^\ast$}
\end{center}
\vskip 0.1truein
\begin{center}
\vspace{6mm}
\bf{Javier Mas and Alexandre Serantes}
\end{center}
\vspace{0.5mm}

\begin{center}\it{
Departamento de  F\'\i sica de Part\'\i  culas \\
Universidade de Santiago de Compostela, \\ 
and \\
Instituto Galego de F\'\i sica de Altas Enerx\'\i as IGFAE\\
E-15782 Santiago de Compostela, Spain}
\end{center}

\begin{abstract}
We study the dynamics of a spherically symmetric thin shell of perfect fluid embedded in $d-$dimensional Anti-de Sitter space-time. In global coordinates, besides collapsing solutions, 
oscillating solutions are found where the shell bounces back and forth between two  radii. The parameter space where  these oscillating solutions exist is scanned in arbitrary number of dimensions. As expected AdS$_3$ appears to be singled out. 
\end{abstract}

\let\thefootnote\relax\footnotetext{Emails: javier.mas@usc.es, alexandre.serantes@gmail.com}
\let\thefootnote\relax\footnotetext{$^{\large\ast}$ Contribution to the proceedings of the ``VII Black Holes Workshop", Aveiro, Portugal, 18-19 December 2014. Preprint of an article published in International Journal of Modern Physics D, Vol. 24, No. 9 (2015) 1542003 © World Scientific Publishing Company DOI: 10.1142/S0218271815420031}

\section{Introduction}

The  non-linearity inherent to Einstein field equations makes the obtention of exact analytical solutions a hazardous program. Among the simplifications that allow to ease this task,  the construction of thin shell space-times stands out as a manageable approximation. Although idealised, thin shells are extremely useful, both from a conceptual and a computational perspective, since they provide nice solvable models in which trademark processes, such as gravitational collapse, can be explicitly addressed\cite{poisson}.
\\\\
On the other hand, solving gravitational problems in Anti-de Sitter space (AdS) has become a theoretical laboratory were one may want to test  results and solutions known from asymptotically flat space-times. Needless to say, this interested has been triggered by the so called AdS/CFT correspondence, also known as holographic gauge/gravity duality\cite{hubeny}.
In such a context, a static black hole in AdS is dual to a thermal state in the CFT that lives in the AdS boundary.  Pushing this identification beyond the static situation, it is natural to conjecture that thermal physics out of equilibrium in CFT can be modelled by dynamical gravitational processes that involve time dependent horizons. For small perturbations this has led to  successful computation of hydrodynamic transport coefficients from quasi-normal modes\cite{Kovtun:2004de}. Large gravitational collapses of some matter field configuration  should provide a dual description of the thermalisation process of a strongly coupled CFT from some out-of-equilibrium initial state\cite{Danielsson:1999fa}. This relaxation process is inaccessible to perturbative field theory techniques, but it is also difficult to treat from the gravitational point of view due to the non-linearity of the partial differential equations involved. 
\\\\
This is the reason why collapsing thin shells in asymptotically AdS spaces have been widely studied in the framework of gauge/gravity duality.
The simplest examples involve flat shells that fall inside planar AdS$_{d+2}$ space, where the dual quantum field theory lives on the  $(d+1)$-dimensional  Minkowskian  boundary. Concerning the kind of matter that makes up the shell, null dust leading to Vaidya space-times has been considered\cite{aal,  bala1}. More general kinds of matter have also been treated recently\cite{knstv1}. 
\\\\
Collapsing shells in global AdS should model  holographically the relaxation of an isolated   quantum system of finite size. Having a boundary conformal to $S_d \times \mathbb{R}$, the natural expectation is that new phenomena may arise, as the sphere introduces a new length scale into the game.
This expectation is reinforced by the fact that, with the help of numerical techniques, exotic behaviours have  been observed in collapses involving a massless scalar field. In the Poincar\'e patch, both a flat shell and a massless scalar pulse lead to direct black hole formation after the first infall towards the Poincar\'e horizon\cite{wu}.  In global coordinates, however, the same matter configuration may exhibit a transient oscillatory behaviour if its mass is sufficiently small. The thick shell of massless scalar field  bounces back and forth between the origin and the AdS boundary several times while it thins out, before ending up forming a black hole\cite{br,bll}. The initial simulations led the authors of Ref.~\cite{br} to conjecture that AdS is non-linearly unstable towards formation of a black hole no matter how small the initial perturbation is. Later on, some numerical evidence was provided to believe that there are ``islands of stability" in the space of initial conditions. The situation at present is not  settled but there are evidences that suggest  that these islands contain each one an exactly periodic solution which governs the stability. In some cases, it has been posible to construct explicitly \cite{rm} such exactly periodic solutions.
\\\\
Our initial intention was to look for a simple analytical model that could reproduce such periodic behaviour. This  led us to consider spherically symmetric $d-$dimensional thin shells embedded in global AdS$_{d+2}$ space-time. Restricting to a family of linear equations of state, we will show that there are  regions in parameter space where the shell undergoes an exactly periodic motion. It is worth stressing  that, in contrast, in  planar AdS shells never bounce back once infalling \cite{aleksi}. 

\section{Shell Dynamics}

The shell world-volume $\Sigma$ is a codimension-1 hypersurface that divides the $(d+2)$-dimensional background space-time $\mathcal{M}$ in two distinct regions:   outside, $\mathcal{M_+}$, and inside, $\mathcal{M_-}$. Due to the spherical symmetry of the problem, we know, by Birkhoff's theorem, that the space-time metric $g$ takes the Schwarzschild-AdS 
form on both $\mathcal{M_+}$, $\mathcal{M_-}$. Choosing standard Schwarzschild coordinates $x_\pm = (t_\pm, r_\pm, \theta_{1}, ..., \theta_{d})$ to cover $\mathcal{M}_\pm$, 
we find that, in this particular coordinate system 
\begin{equation} 
ds^2_\pm = - f_\pm(r_\pm) dt_\pm^2 + f_\pm(r_\pm)^{-1}dr_\pm^2 + r_\pm^2 d\Omega_{d}^2 \label{g} \\ 
\end{equation}
where
\begin{equation}
f_\pm(r_\pm) = 1 + \frac{r_\pm^2}{l_\pm} - \frac{m_\pm}{r_\pm^{d-2}} ~ .
\end{equation}
As usual, $d\Omega_d^2$ is the metric of a unit round $d$-dimensional sphere, and the AdS radius $l$ is related to the cosmological constant $\Lambda = -\frac{d(d+1)}{2 l^2}$. In what follows, we will restrict ourselves to the case where the shell has no influence on the cosmological constant, so we are going to set $l_+ = l_- = l = 1$ by an appropriate  choice of units. Furthermore, we  assume that the space-time inside the shell is empty AdS$_{d+2}$ and hence fix $m_- = 0$. In such case, $m_+ \equiv  m$ sets the total ADM mass of the system.
\\\\
\noindent Let the shell world-volume  $\Sigma$ be parameterized with coordinates $y = (\tau, \theta_1, ..., \theta_d) \label{y}$, where $\tau$ is the proper time of an comoving observer. The shell embedding in the ambient space-time is given parametrically by the function 
\begin{equation} x_s(y) = (t_{\pm,s}(\tau), r_{\pm,s}(\tau), \theta_1, ..., \theta_d)~. \label{embedding} \end{equation}
The tangent space $\mathcal{T}_p \mathcal{M}$ of any point $p \in \Sigma$ admits a basis formed by $d+1$ vectors $e_a = e^\alpha_a \partial_{x^\alpha}$, tangent to $\Sigma$, and one vector $n = n^\alpha \partial_{x^\alpha}$, orthogonal to $\Sigma$. Explicitly, 
\begin{eqnarray} 
e_{\tau, \pm} &=&   \dot{t}_{\pm,s} \partial_{t_\pm} + \dot{r}_{\pm,s} \partial_{r_\pm} \\ 
e_{\theta_i, \pm} &=&   \partial_{\theta_i} \\
n_{\pm} &=&  + \left(f^{-1}_{\pm,s} \dot{r}_{\pm,s} \partial_{t_\pm}  + f_{\pm, s} \dot{t}_{\pm, s} \partial_{r_\pm} \right) \label{n}
\end{eqnarray}
The overall positive sign of $n_\pm$ is fixed by requiring that $n_\pm$ is always directed from $\mathcal{M_-}$ to $\mathcal{M_+}$\cite{poisson}. 
\\\\
The embedding \eqref{embedding} is not arbitrary: in order for the whole space-time $\mathcal{M}$ to solve Einstein equations, the so called {\it Israel junction conditions} must be satisfied (see Ref.~\cite{poisson}). The first junction condition states that the  induced metric $h_{ab}$ on $\Sigma$ must be continuous across $\Sigma$
\begin{equation} \left[h_{ab} \right] = 0 \label{israel1} \end{equation}
where the brackets stand for the jump.
The second junction condition relates the jump of the extrinsic curvature $K_{ab}$ with the matter composition of the shell,  
\begin{equation} \left[K_{ab} - h_{ab} K \right] = - 8 \pi G S_{ab} = - S_{ab} \label{israel2} \end{equation}
where $K \equiv h^{ab} K_{ab}$, $S_{ab}$ is the shell energy-momentum tensor and we have chosen units such that $8 \pi G = 1$. 
Projecting  $g$ onto $\Sigma$ to find the induced metric $h_{ab} = g_{\alpha \beta} e^\alpha_a e^\beta_b$ we get 
\begin{equation}dh^2_\pm = h_{\pm ab} dy^a dy^b =  \left( - f_{\pm,s} \dot{t}_{\pm,s}^2 + f_{\pm,s}^{-1} \dot{r}_{\pm,s}^2  \right) d\tau^2 + r_{\pm,s}^2 d\Omega_{d}^2 ~. \label{h} \end{equation}
The choice of $\tau$ as comoving time fixes $h_{\tau \tau} = -1$, whence it follows that 
\begin{equation}  
\dot{t}_{\pm,s} =  \frac{\beta_\pm}{f_{\pm,s}} \label{tdot} 
\end{equation}
with 
\begin{equation}
\beta_\pm = \sqrt{f_{\pm,s} + \dot{r}_{\pm,s}^2}~.\label{beta}
\end{equation}
We have taken the positive root of $\dot{t}_{\pm,s}$, as we want the shell trajectory to be future oriented. Equation \eqref{tdot} together \eqref{beta} accomplishes two tasks.  It  ensures that the $\tau\tau$ component of the first junction condition \eqref{israel1} is  satisfied, and gives the correct normalisation to the vector $n$ in \eqref{n}, $n^2=1$. It also implies that it is imposible to cover the entire space-time $\mathcal{M}$ with a globally defined time-like Schwarzschild coordinate, as the embedding functions $t_{\pm,s}(\tau)$ will differ at the shell. On the other hand, the radial coordinate $r_\pm$ has to be continuous  since $r_{+,s}(\tau) = r_{-,s}(\tau) \equiv r_s(\tau)$ must hold to signal unambiguously the shell's radial position. This condition, together with \eqref{tdot}, ensures that all components of \eqref{israel1} are satisfied. From now on, we take these facts into account and change correspondingly our $\mathcal{M_\pm}$ coordinate system to $x_\pm = (t_\pm, r, \theta_i)$. \\\\
The extrisic curvature  is the pullback of the  Lie derivative of the ambient metric $g$ along $n$. Several equivalent expressions can be found in the literature\cite{poisson}
\begin{equation} K_{ab} = \frac{1}{2} e^\alpha_a e^\beta_b \left(\mathsterling_n g \right)_{\alpha \beta} = e^\alpha_a e^\beta_b \nabla_\alpha n_{\beta} = - n_\mu \left( \frac{\partial x^\mu_s}{\partial_{y^a} \partial_{y^b}} + \Gamma^\mu_{\alpha \beta} e^\alpha_a e^\beta_b \right)\end{equation}
where the orthogonality condition $e^\alpha_a n_\alpha = 0$ is used. In our particular  setup \eqref{h}, its non-zero components and trace are 
\begin{equation} K^\tau_{\pm,\tau} = \frac{\dot{\beta}_\pm}{\dot{r}_s} ~~~~~~~ K^{\theta_i}_{\pm,\theta_i} = \frac{\beta_\pm}{r_s}  ~~~~~~~  K =  \frac{\dot{\beta}_\pm}{\dot{r}_s} + d \frac{\beta_\pm}{r_s}\end{equation}
The diagonal nature of $K^a_b$, together with the second Israel junction condition \eqref{israel2}, leave little room  for the form of the shell stress-energy tensor $S^a_b$, which must be of the  perfect fluid form 
\begin{equation} S^a_b = {\rm diag}(- \sigma, p, ..., p)\end{equation}
where $\sigma$ will be  the shell energy density and $p$ the shell pressure. Due to spherical symmetry, $p$ is independent of the particular angular direction considered. In components, \eqref{israel2} reads now 
\begin{eqnarray} 
\left[K^{\theta_i}_{\theta_i} \right] = \frac{\left[\beta \right]}{r_s} &=& -\frac{1}{d}\sigma \label{israel2component1} \\
\left[K^\tau_\tau \right] = \frac{\left[\dot{\beta} \right]}{\dot{r}_s} &=& p + \frac{d-1}{d} \sigma \label{israel2component2}
\end{eqnarray}
With our choices for $f_\pm$, we always have  $\beta_+ \leq \beta_-$ and, therefore, $\sigma \geq 0$. As usual, equations \eqref{israel2component1}, \eqref{israel2component2} need to be supplemented with an equation of state which relates the shell energy density and pressure.  
At this point, we introduce a simplification by restricting our analysis to the case where this equation of state is linear. In AdS$_{d+2}$ we shall write
\begin{equation} p = \frac{\alpha}{d}\, \sigma ~. \label{eos} \end{equation}
The parameter $\alpha$ determines the kind of matter the shell is made of. Taking $\alpha \in \left[0, 1 \right]$ it interpolates between dust ($\alpha = 0$) and conformal matter ($\alpha = 1$). This choice of equation of state, together with the positivity of $\sigma$, implies that $\sigma + p \geq 0$, so that the weak energy condition is respected. 
With the choice \eqref{eos}, equations \eqref{israel2component1}, \eqref{israel2component2} can be solved explicitly. The final result is  that the shell dynamics is fully equivalent to the one-dimensional motion of a particle in an effective potential $V_{eff}$
\begin{equation} 
\dot{r}_s^2 + V_{eff} = 0 \label{eom2} 
\end{equation}
where 
\begin{equation}
V_{eff} = 1 + r_s^2 - \frac{1}{2}m r_s^{1-d} - \frac{m^2}{4 M^2} r_s^{2\alpha}- \frac{1}{4}M^2r_s^{-2\left(d-1+\alpha \right)}\, . \label{V}
\end{equation}
 $M$ is an integration constant that sets the shell's proper energy $E$, defined as $E \equiv vol(S_d) r_s^d \sigma = vol(S_d) d \,  r_s^{-\alpha}M$, where $vol(S_d)$ is the volume of the $d$-dimensional unit sphere. \\\\
Notice that the potential $V_{eff}$ is invariant under $m \rightarrow m$, $M \rightarrow \frac{m}{M}$ and $\alpha \rightarrow - (d - 1 +\alpha)$ so any result we may obtain is also going to hold in the range $\alpha \in \left[-d, -d+1 \right]$, modulo the appropriate $M$ redefinition. The weak energy condition will be still satisfied. 

\section{Oscillating Solutions}

From \eqref{eom2}, we know that the region where the shell is allowed to move is the -possibly disconnected- set of radial intervals for which $V_{eff} \leq 0$. The asymptotic behaviour of $V_{eff}$ is as follows:
\begin{itemize}
\item $V_{eff} \rightarrow -\infty$ as $r \rightarrow 0$, with $V_{eff} \sim - 1/4 M^2 r^{-2(d-1 +\alpha)}$ for $r \ll 1$\\
\item $V_{eff} \rightarrow \infty$ as $r \rightarrow \infty$, with $V_{eff} \sim r^2$ for $r \gg 1$
\end{itemize}
Oscillating shell trajectories, if they exist, are confined to an intermediate radial region where the potential $V_{eff}$ develops a well. The turning points are given by two radii $r_{\pm}$ where $V_{eff}(r_{\pm})=0$ (subindices here do not refer to the inner or outer regions to the shell).
The goal now is, fixing $d$, $\alpha$ and the space-time ADM mass $m$, find in what range of $M$ oscillating solutions appear. It turns out that this $M$ region is bounded by two shell rest energies  
$M_{l, u}$, with $M_l \leq M_u$, such that
\begin{itemize}
\item For $M = M_u$ there is a local minimum of $V_{eff}$ touching the $V_{eff} = 0$ axis. This corresponds to a shell in equilibrium. 
\item For $M = M_l$ there is a local maximum of $V_{eff}$ touching the $V_{eff} = 0$ axis. This corresponds to the transition between oscillatory and collapsing behaviour. 
\end{itemize}

\begin{figure}[h!]
\begin{center}
\includegraphics[width=10cm]{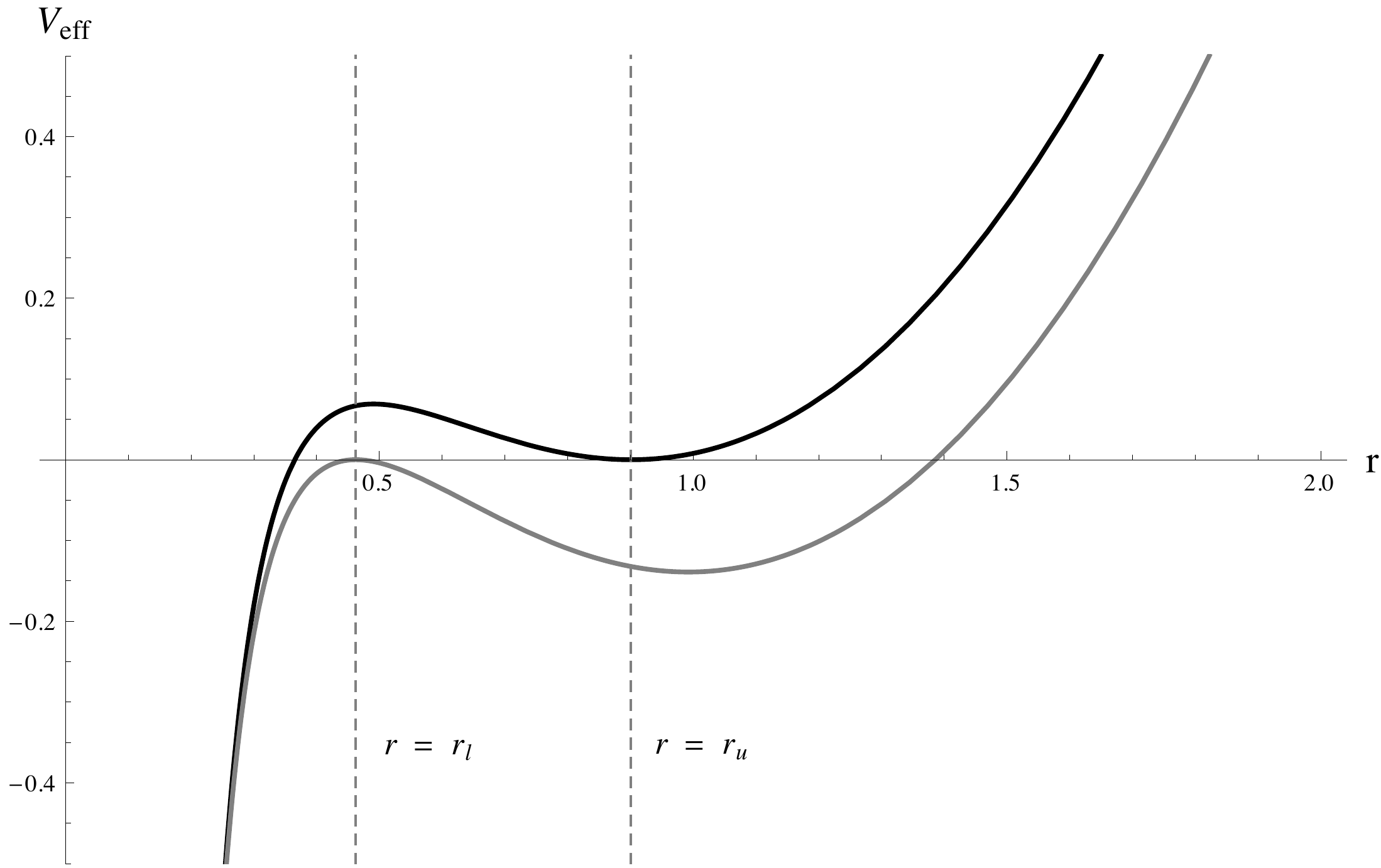}
\end{center}
\caption{\label{1} {\em Typical form of the potential $V_{eff}$ in the limiting cases that bound the oscillating shell existence region. In black, the potential at $M = M_u$. In gray, the potential at $M = M_l$. Dashed gray lines correspond to $r_l$ and $r_u$. The parameter values are $d=3$, $\alpha= 0.5$, $m=0.1$, $M_u = 0.0359$, $M_l = 0.0346$, $r_l = 0.426$, $r_u = 0.901$.}}
\end{figure}

\noindent Fig. \ref{1} depicts both limiting situations on AdS$_5$. To find the values of $M_{l, u}$, we have to solve the system of equations given by $V_{eff} = \partial_r V_{eff} = 0$. The solution is easily obtained in implicit form by taking $m = m(d,\alpha, r)$, $M = M(d,\alpha, r)$, which will be called the {\it existence curves}
\begin{eqnarray}
m(d, \alpha, r) &=& \frac{4 r^{d-1}(\alpha - (1-\alpha)r^2)(d-1+ \alpha + (d+\alpha)r^2)}{(d-1+2\alpha)^2\left(1+r^2\right)}\label{msol}\\
M(d, \alpha, r) &=& \frac{2r^{d-1+\alpha}(\alpha - (1-\alpha)r^2)}{(d-1+2\alpha) \sqrt{1+r^2}} \label{Msol}
\end{eqnarray}
As the $d= 1$ existence curves display peculiar  properties we will discuss this case separately. 

\subsection{Oscillating shells in $d > 1$}

\begin{figure}[h]
\begin{center}
\includegraphics[width=15cm]{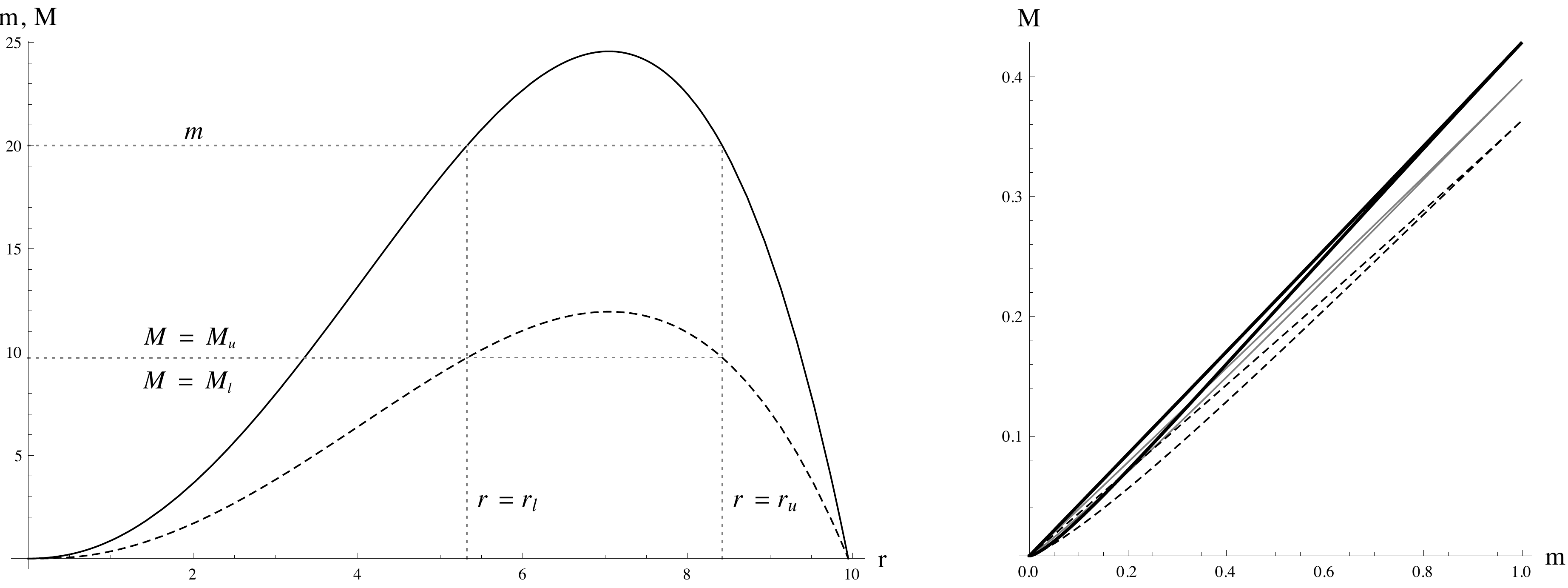}
\end{center}
\caption{\label{2} {\em Left: existence curves for oscillating solutions in $d=3$ and $\alpha = 0.99$. In solid black, $m(d,\alpha, r)$. In dashed black, $M(d,\alpha, r)$. The gray dashed curves illustrate the construction described in the text. Notice that the range of M is so tiny that the two lines depicting $M_l$ and $M_u$ look coincident. Right: $M(3, \alpha, m)$ for $\alpha = 0.2, 0.5, 0.9$ -gray, black dashed and black curves respectively-. Any point within the region bounded by each curve corresponds to a particular oscillating solution. All quantities are measured in units of $m_{max}(d, \alpha)$. }}
\end{figure}

\noindent The existence curves allow a straightforward computation of $M_l$ and $M_u$. The procedure is illustrated in Fig. \ref{2} (left). First, we choose some $m$ and solve numerically the equation $m = m(d, \alpha, r)$ for $r$ at fixed $d$ and $\alpha$. The output are two radii, $r_l$ and $r_u$, such that $r_l < r_u$: $r_u$ signals the position of the axis-touching minimum of $V_{eff}(d,\alpha,m,M_u)$, while $r_l$ signals the position of the axis-touching maximum of  $V_{eff}(d,\alpha,m,M_l)$ -see Fig.~\ref{1}-. Inserting $r_{l, u}$ into equation \eqref{Msol} gives back the numerical values of $M_{l, u}$ and fixes completely the form the potential $V_{eff}$. For any $M \in (M_l, M_u)$, it is guaranteed that $V_{eff}$ possesses an oscillating solution.\\\\
\noindent Looking at Fig. \ref{2} (left), it is neatly seen that, at fixed $d$ and $\alpha$, there is a maximum mass, $m_{max}(d,\alpha)$, above which the construction just described can not be performed. Therefore, oscillating shell trajectories only exist for sufficiently light shells; above $m_{max}(d, \alpha)$ there are only collapsing solutions. The fact that there are no oscillating geometries when the mass of the system surpasses a certain threshold is a feature that this simple model shares with more realistic setups, for instance, a minimally coupled massless scalar field in global AdS\cite{br,bll}. 

\begin{figure}[h!]
\begin{center}
\includegraphics[width=10cm]{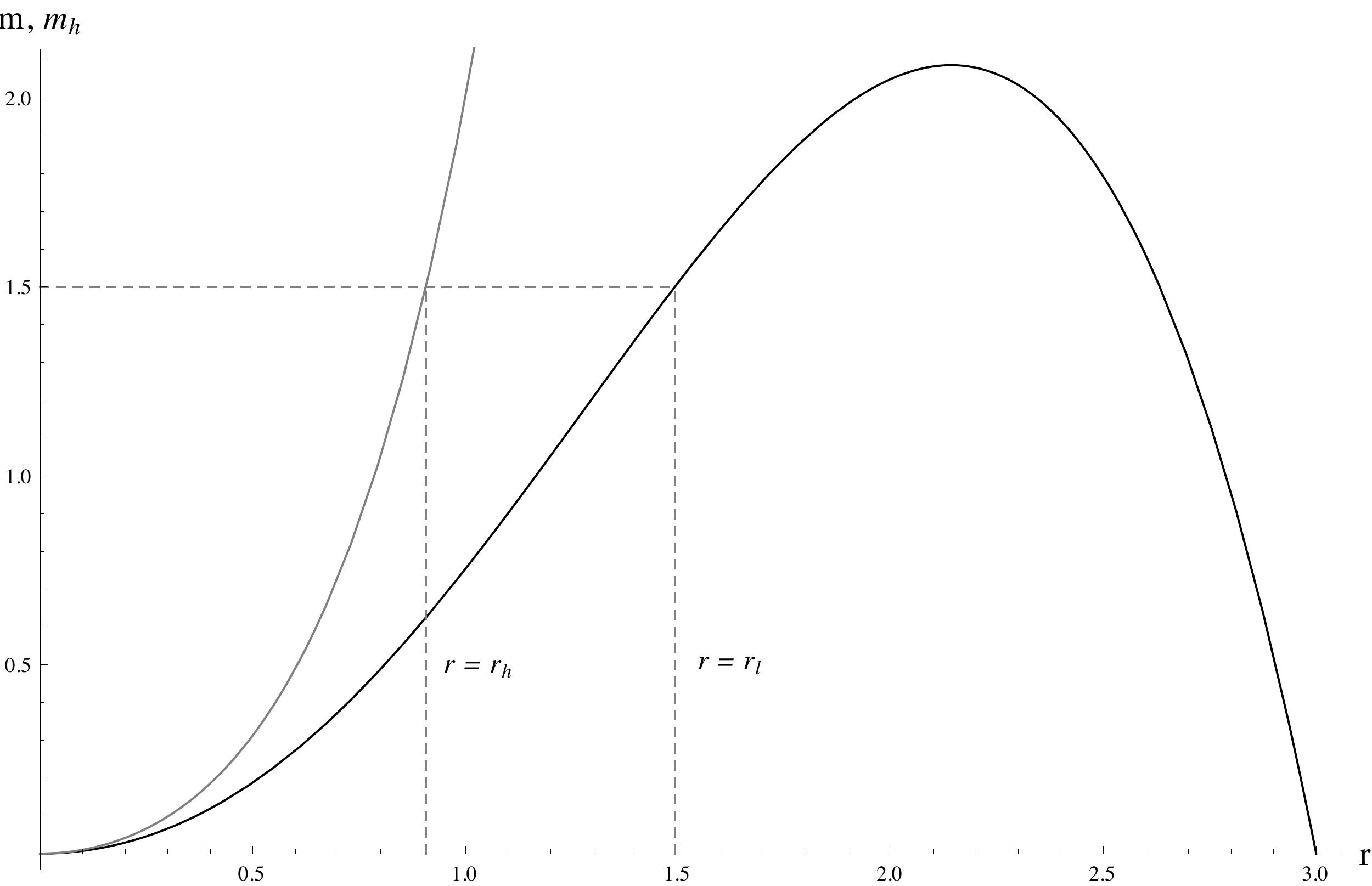}
\end{center}
\caption{\label{eh} {\em Position of the event horizon -gray- versus $m$ existence curve -black- in $d=3$ for $\alpha = 0.9$. The line $m = 1.5$ is drawn -gray dashed-, as well as its intersection points with $m$, $r_l$ and $m_h$, $r_h$ -gray dashed-}}
\end{figure}

\noindent Concerning $M$, it is important to stress that its value sets an energy scale that is independent of $m$. Tuning $M$ deforms $V_{eff}$ and, in particular, shifts the upper turning point $r_+$. Hence, one can think of this parameter in terms of the initial radius where the shell is released from rest and starts falling and, in consequence, $M$ would be related to its potential energy.\\\\
\noindent In Fig. \ref{2} (right) we offer several parametric plots of $M(d, \alpha, r)$ as function of $m(d,\alpha,r)$. In each curve, the upper branch corresponds to $M_u$, while the lower branch corresponds to $M_l$. In the $m \rightarrow 0$ limit, the upper $M_u$ branch asymptotes to a line,  
\begin{equation} M_u(\alpha, m) \sim \frac{1}{2} \left(1 - \alpha \right)^{\frac{1-\alpha}{2}}\alpha^{\alpha/2} m~. \end{equation}
\noindent Oscillating solutions exist for $M$ in a narrow window around this upper branch of static shells. This window closes at $m_{max}(d, \alpha)$. If the two scales set by $m$ and $M$ were not independent, this finite-size region where oscillating solutions reside would degenerate into a line and they would cease to exist.\\\\
\noindent To start scanning the behavior of the existence region with respect to $\alpha$, let us note that both $m(d, \alpha, r)$ and $M(d, \alpha, r)$ attain their maxima at the same radius
\begin{equation} r_{max}(d, \alpha)=\frac{1}{\sqrt{2}}\frac{\sqrt{2\alpha(d-1+\alpha)-(d+1) + \sqrt{(d+1)^2 - 4(d-1)\alpha - 4\alpha^2)}}}{\sqrt{(1-\alpha)(d+\alpha)}}\, . \label{rmax}\end{equation}
which also controls $m_{max}(d,\alpha)$, 
\begin{equation}  m_{max}(d, \alpha) = 4 r_{max}(d,\alpha)^{d+1}\frac{ \sqrt{(d+1)^2 - 4(d-1)\alpha - 4\alpha^2} - (d-1) }{(d-1+2\alpha)^2} \, . \label{mmax}\end{equation}
In the conformal fluid  limit $\alpha \rightarrow 1$, $r_{max}(d, \alpha)$ diverges as $r_{max}(d, \alpha) \sim \sqrt{\frac{d-1}{d+1}} \frac{1}{\sqrt{1-\alpha}}$, so we expect $m_{max}(d, \alpha)$ 
to be also divergent. In fact 
\begin{equation}  m_{max}(d, \alpha) \sim \frac{8}{d^2-1} \left(\frac{d-1}{d+1}\right)^{\frac{1+d}{2}} (1- \alpha)^{\frac{1-d}{2}}~.\label{mdiver}
\end{equation}
\noindent This result shows that the maximum mass for which oscillating solutions exist grows unbounded as $\alpha \rightarrow 1$; for $d > 1$, there are oscillating solutions with arbitrary high ADM energy as long as the shell matter is sufficiently near conformallity. In the opposite limit (pressure-less dust) $\alpha \rightarrow 0$
\begin{equation}  m_{max}(d, \alpha) \sim \frac{8}{(d-1)^2} \left(\frac{d-1}{d+1} \right)^{\frac{1+d}{2}} \alpha^\frac{1+d}{2} \end{equation}
and the allowed region for oscillating trajectories shrinks down to zero. Physically this means that there are no pressure-less oscillating solutions, {\em i.e.}
in order to be stable against gravitational collapse, the shell matter must have some self-interaction. As an aside, note that the limiting behaviour of $M(d, \alpha, r)$ as 
$\alpha \rightarrow 0, 1$ just follows the behaviour of $m(d, \alpha, r)$, since 
\begin{equation}  M(d, \alpha, r) = r^\alpha\frac{(d-1+2\alpha) \sqrt{1+ r^2}}{2 (d-1+\alpha + (d+\alpha)r^2)} m(d, \alpha, r)~. \label{Mmrelation}\end{equation}
\noindent An important consistency check is to verify that $r_-$ lies outside the position of the event horizon, $r_h$. Otherwise, instead of having an oscillating solution, the shell trajectory would represent direct gravitational collapse starting from $r_+$.  As $r_- \geq r_l$, it is sufficient to show that $r_l \geq r_h$.  The static event horizon location, 
$r_h(d, m)$, is the solution of the equation $1+ r_h^2 - m r_h^{1-d} = 0$. Since we don't know explicitly $r_l = r_l(d, \alpha, m)$, but instead $m = m(d, \alpha, r_l)$, we are going to define 
correspondingly $m_h(d, r) \equiv r^{d-1}(1+r^2)$. It is easy to see that, if $m_h(d, r) - m(d, \alpha, r) \geq 0$ for all $d, \alpha, r$, the consistency condition $r_l \geq r_h$ always holds 
-see Fig. \ref{eh}-. We find that 
\begin{equation}  m_h(d, r) - m(d, \alpha, r)  = r^{d-1} \frac{\left(d-1 + (d+1)r^2\right)^2}{\left(1+r^2 \right)\left(d-1+2 \alpha \right)^2} \label{outside} \end{equation}
 which is positive definite. Thus, every oscillating shell trajectory found lies entirely outside the event horizon of the would-be black hole. 
\medskip 

\subsection{Oscillating shells in d = 1}

\begin{figure}[h!]
\begin{center}
\includegraphics[width=13cm]{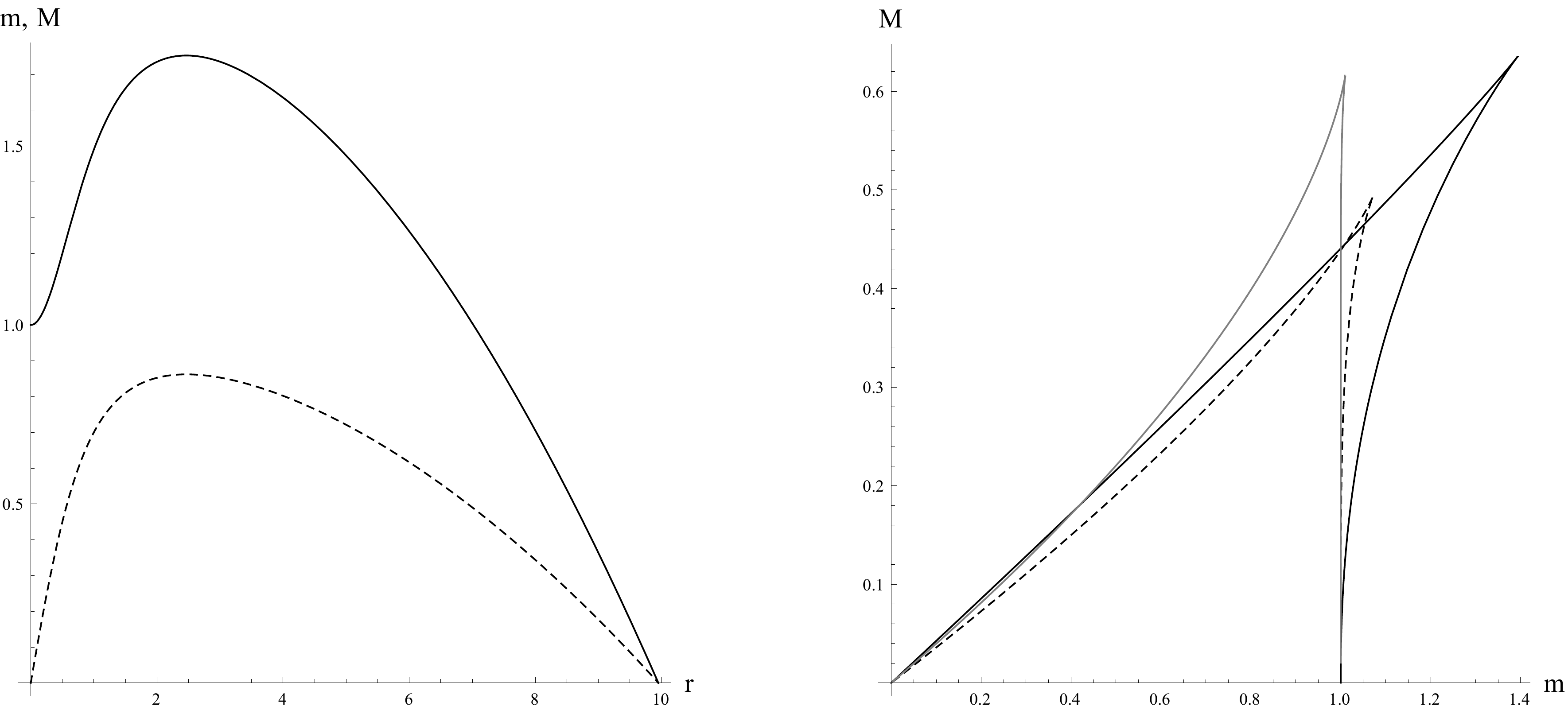}
\end{center}
\caption{\label{ads3} {\em Left: existence curves for oscillating  solutions in $d=1$ and $\alpha = 0.99$. In solid black, $m$ as a function of $r$. In dashed black, $M$. Right: $M(1, \alpha, m)$ for $\alpha = 0.2, 0.5, 0.9$ -gray, black dashed and black curves respectively-.}}
\end{figure}

The case $d = 1$ is special because, as mentioned before, some general properties seen in $d > 1$ do not hold anymore. Representative existence curves are plotted in Fig. \ref{ads3}. The analytic form of $m(d, \alpha, r), M(d, \alpha, r)$ for $d = 1$ is 
\begin{eqnarray}
m(1, \alpha, r) \equiv m(\alpha, r) &=& 1+r^2 - \frac{r^4}{(1+r^2)\alpha^2}\label{msol3}\\
M(1, \alpha, r) \equiv M(\alpha, r) &=&  r^\alpha \frac{\alpha - (1- \alpha)r^2}{\alpha \sqrt{1+r^2}}  \label{Msol3}
\end{eqnarray}

\begin{figure}[h!]
\begin{center}
\includegraphics[width=15cm]{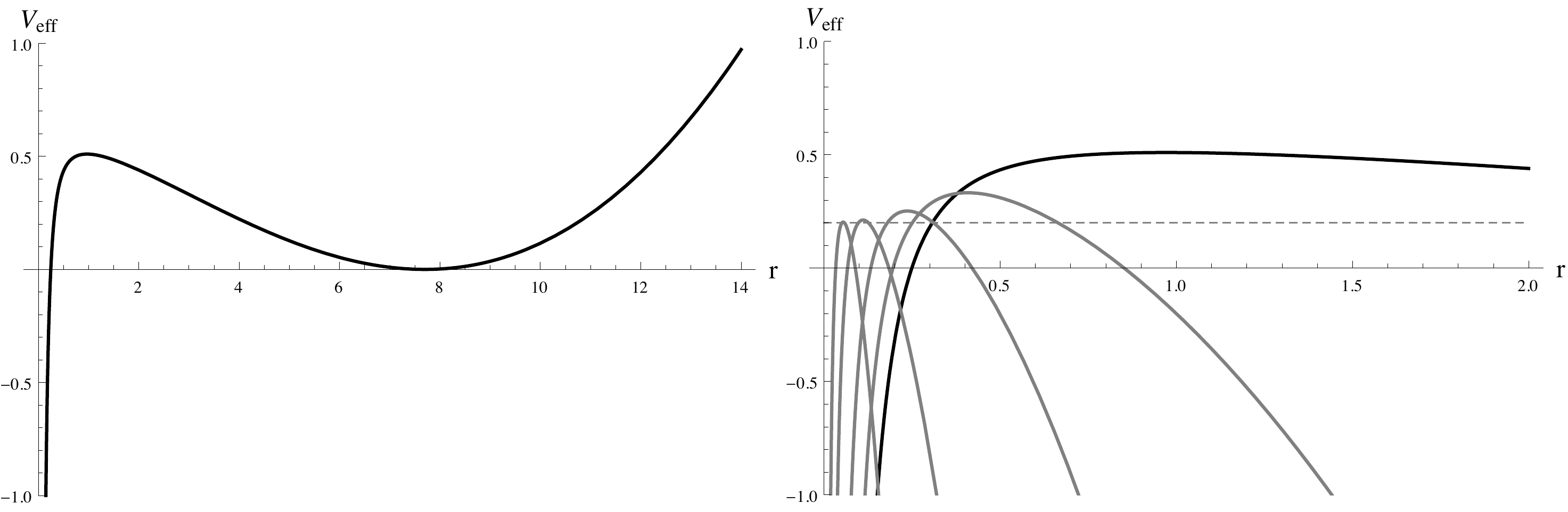}
\end{center}
\caption{\label{barrier} {\em Left: for $\alpha = 0.99$ and $m = 0.8$, potential $V_{eff}$ at $M = 0.3899$, which corresponds to a static shell located at $r_u = 7.699$. Right: evolution of the potential barrier for the oscillating shell at different $M$. The black curve corresponds to $V_{eff}$ on the left picture. From left to right, gray curves represent potentials with $M = 0.3$, $M = 0.2$, $M =0.1$ and $M = 0.05$. Note that $r_{barrier}$ gets smaller as $M$ does, while $V_{barrier} \rightarrow 1 - m = 0.2$ in this case -black dashed-}}
\end{figure}

\noindent It is easy to prove that $r_l \geq r_h$, as equation \eqref{outside} still holds. The major difference with respect to the $d > 1$ case comes when we evaluating the allowed mass range for oscillating shell trajectories. At $d = 1$
\begin{equation} r_{max}(\alpha) = \left(\frac{1+\sqrt{1-\alpha^2}}{\alpha^2} - 1\right)^{-\frac{1}{2}} \end{equation}
is still divergent as $\alpha \rightarrow 1$, although in a  milder way, $r_{max}(\alpha) \sim (1- \alpha)^{-\frac{1}{4}}$ instead of the $(1-\alpha)^{-\frac{1}{2}}$ divergence in \eqref{rmax}. However, 
\begin{equation} m_{max}(\alpha) = \frac{2}{1+\sqrt{1-\alpha^2}} \end{equation}
so, unlike the $d>1$ case, for $d = 1$ the maximum allowed mass of an oscillating solution does not grow without bound as $\alpha \rightarrow 1$. Instead, it goes to a finite value, $m = 2$ (in our conventions). Note also that, in the $\alpha \rightarrow 0$ limit, the $m$-range  does not close down. Instead, $m_{max}(\alpha) \rightarrow 1$ as $\alpha \rightarrow 0$. In AdS$_3$ there are oscillating shell solutions for any $m \leq 1$, even for dust made shells.\\\\
\noindent Looking at figure Fig. \ref{ads3}, we observe that there is a qualitative difference between shells of mass above and below the threshold  $m = 1$.  Each mass $m$ belonging to the interval $[1, m_{max}(\alpha)]$ has two associated radii, $r_l$ and $r_u$, such that, like in higher dimensions, $r_l$ signals the position of the $V_{eff}=0$ axis-touching maximum, while $r_u$ signals the position of the axis-touching minimum. However, for $m < 1$ the axis-touching maximum disappears. This does not mean that shells with $m < 1$ are allowed to reach the point $r = 0$, because the potential barrier does not vanish at any finite $M$: as $M \rightarrow 0$, the maximum of the barrier gets radially displaced towards the origin and tends to a constant value. At the same time, $r_+$ grows unbounded. In the same way that oscillating shells can not cross their Schwarzschild radius, in AdS$_3$ there is also a mechanism forbidding the possibility of reaching $r = 0$: the shells can not form naked singularities. Again this is in parallel with the phenomenology shown by a massless scalar pulse in AdS$_3$\cite{bj, j}. The behaviour of the barrier is depicted in figure Fig. \ref{barrier}.\\\\
\noindent It is easy to explain the behaviour of $V_{eff}$ as $M \rightarrow 0$ in analytic terms. Let $r_{barrier}$ be the position of the maximum of the potential barrier. If $r \ll1$, $\partial_r V_{eff} \sim \alpha\left(M^2 r^{-2\alpha} - \frac{m^2}{M^2}r^{2\alpha}\right)$, so \begin{equation} r_{barrier} \sim \left(\frac{M}{\sqrt{m}}\right)^{\frac{1}{\alpha}}~.\end{equation}
Therefore, $r_{barrier} \rightarrow 0$ as $M \rightarrow 0$. At $r=r_{barrier}$, the potential is finite 
\begin{equation} V_{barrier} \equiv V_{eff}(r_{barrier}) \sim 1 - m~. \end{equation}
Regarding the large $r$ turning point, $r_+$, note that if $V$ has a root $r_+$ such that $r_+ \gg 1$, we have that $V(r_+) \sim r_+^2 - \frac{m^2}{4 M^2} r_+^{2\alpha} = 0$, which is solved by 
\begin{equation} r_+ = \left( \frac{m}{2 M} \right)^{\frac{1}{1-\alpha}}~. \end{equation}
This proves that $r_+ \rightarrow \infty$ as $M \rightarrow 0$.

\section{Conclusions}

We have explored a family of solutions of General Relativity in global AdS exhibiting periodic behaviour. It involves thin spherical shells of matter governed by linear equations of state  that range from dust to conformal fluids. We have provided exact analytic expressions for the regions of existence of such solutions and analysed their behaviour in certain limits. The conclusions match in a natural way with similar results obtained in numerical simulations of realistic shells of massless scalar field. In particular, we have checked that oscillating shells never reach the Schwarzschild horizon.  The case of AdS$_3$ has been separately analysed due to the presence of the mass gap. 
\\\\
 As several simplifying assumptions have been made, each one  can be relaxed independently in order to test the robustness of the results obtained here. For instance, more general equations of state, {\it i.e.} polytropes, can be considered. The requirement that the interior space-time $\mathcal{M}_-$ is an empty AdS space with the same cosmological constant as the exterior one can also be lifted. Finally going beyond, spherical symmetry is a natural extension of this work. While the case for rotating oscillating solutions looks feasible, the addition of angular deformations appears as a formidable challenge.
 \\\\ 
In the light of gauge/gravity duality, these oscillating geometries should correspond to out-of-equilibrium field theory states that never thermalise\cite{adslms}. As has been stressed, the simple nature of the solutions found would allow a relatively easy calculation of holographic proxies of field theory quantities (such as entanglement entropy\cite{knstv1}, two-point correlation functions or mutual information) that would help to characterise the non-thermalising state. This issue is currently under research. 

\section*{Acknownlegments}

We want to thank Oscar Dias,  Roberto Emparan, Ville Ker\"anen and Olli Taanila for discussions.
The work of J.M.  is supported in part by the spanish grant  FPA2011-22594,  
by Xunta de Galicia (GRC2013-024), by the  Consolider-CPAN (CSD2007-00042), 
and by FEDER. A.S. is supported by the European Research Council grant HotLHC ERC-2011-StG-279579 
and by Xunta de Galicia (Conselleria de Educaci\'on).


\begin{thebibliography}{00}       

\bibitem{poisson} 
E. Poisson, {\it A Relativist's Toolkit: 
The Mathematics of Black-Hole Mechanics}, 1st edn. (Cambridge University Press, 2007)


\bibitem{hubeny}
  V.~E.~Hubeny,
  arXiv:1501.00007 [gr-qc].

\bibitem{Kovtun:2004de} 
  P.~Kovtun, D.~T.~Son and A.~O.~Starinets,
  Phys.\ Rev.\ Lett.\  {\bf 94}, 111601 (2005)
  [hep-th/0405231].
  
\bibitem{Danielsson:1999fa} 
  U.~H.~Danielsson, E.~Keski-Vakkuri and M.~Kruczenski,
  JHEP {\bf 0002}, 039 (2000)
  [hep-th/9912209].
 
  
\bibitem{aal}
  J.~Abajo-Arrastia, J.~Aparicio and E.~Lopez,
  JHEP {\bf 1011}, 149 (2010)
  [arXiv:1006.4090 [hep-th]].
   
\bibitem{bala1} 
  V.~Balasubramanian, A.~Bernamonti, J.~de Boer, N.~Copland, B.~Craps, E.~Keski-Vakkuri, B.~Muller and A.~Schafer {\it et al.},
  Phys.\ Rev.\ D {\bf 84}, 026010 (2011)
  [arXiv:1103.2683 [hep-th]].

\bibitem{knstv1}    
V.~Keranen, H.~Nishimura, S.~Stricker, O.~Taanila and A.~Vuorinen,
  Phys.\ Rev.\ D {\bf 90}, no. 6, 064033 (2014)
  [arXiv:1405.7015 [hep-th]].
\\
  arXiv:1502.01277 [hep-th].

\bibitem{wu}   
B.~Wu,
  JHEP {\bf 1210}, 133 (2012)
  [arXiv:1208.1393 [hep-th]].


\bibitem{br}
  P.~Bizon and A.~Rostworowski,
  Phys.\ Rev.\ Lett.\  {\bf 107}, 031102 (2011)
  [arXiv:1104.3702 [gr-qc]].

\bibitem{bll}
  A.~Buchel, L.~Lehner and S.~L.~Liebling,
  Phys.\ Rev.\ D {\bf 86}, 123011 (2012)
  [arXiv:1210.0890 [gr-qc]].

\bibitem{rm}
  M.~Maliborski and A.~Rostworowski,
  Phys.\ Rev.\ Lett.\  {\bf 111}, no. 5, 051102 (2013)
  [arXiv:1303.3186 [gr-qc]].

\bibitem{aleksi} 
V. Keranen, H. Nishimura, S. Stricker, O. Taanila and A. Vuorinen, unpublished notes. 


\bibitem{bj}
  P.~Bizon and J.~Jalmuzna,
  Phys.\ Rev.\ Lett.\  {\bf 111}, no. 4, 041102 (2013)
  [arXiv:1306.0317 [gr-qc]].

\bibitem{j}
  J.~Jalmuzna,
  arXiv:1311.7409 [gr-qc].

\bibitem{adslms}
  J.~Abajo-Arrastia, E.~da Silva, E.~Lopez, J.~Mas and A.~Serantes,
  JHEP {\bf 1405}, 126 (2014)
  [arXiv:1403.2632 [hep-th]].

\end{thebibliography}
\end{document}